\documentclass[aps,prl,superscriptaddress,groupedaddress,nofootinbib,twocolumn]{revtex4}

\usepackage{amsmath,amssymb,amsfonts}
\usepackage[colorlinks]{hyperref}
\usepackage{graphicx}
\graphicspath{{./}}
\newcommand{\draftnote}[1]{{#1}}
\newcommand{\revision}[1]{{#1}}

\newcommand{\bw}{\begin{widetext}}
\newcommand{\ew}{\end{widetext}}
\newcommand{\be}{\begin{equation}}
\newcommand{\en}{\end{equation}}
\newcommand{\bee}{\begin{equation}}
\newcommand{\ene}{\end{equation}}
\newcommand{\bea}{\begin{eqnarray}}
\newcommand{\ena}{\end{eqnarray}}
\newcommand{\bes}{\begin{subequations}}
\newcommand{\ens}{\end{subequations}}
\newcommand{\bef}{\begin{figure}}
\newcommand{\enf}{\end{figure}}
\newcommand{\eq}[1]{Eq.~(\ref{#1})}

\newcommand{\mum}{\mu{\rm m}}
\newcommand{\mm}{{\rm mm}}

\newcommand{\iref}[1]{(\ref{#1})}

\def\thefootnote{\fnsymbol{footnote}}
\def\etal{{\it et al.}}
\def\nb{\, {\rm nb}}

\def\pb{\, {\rm pb}}
\def\ab{\, {\rm ab}}
\def\met{p_T \hspace*{-1.1em}/\hspace*{0.5em}}

\def\met{\vec{{p\!\!\!\slash}}_{T} }

\def\to{\rightarrow}

\def\tev{{\rm TeV}}
\def\gev{{\rm GeV}}

\begin{document}


\title{Probing CP violation in $h\to \tau^{-}\tau^{+}$ at the LHC}

\author{Kaoru Hagiwara}
\email[Electronic address: ]{kaoru.hagiwara@kek.jp}
\affiliation{KEK Theory Center and Sokendai, Tsukuba, Ibaraki 305-0801, Japan}
\affiliation{Department of Physics, University of Wisconsin-Madison, Madison, WI 53706, USA}

\author{Kai Ma}
\email[Electronic address: ]{makainca@yeah.net}
\affiliation{School of Physics Science, Shaanxi University of Technology, Hanzhong 723000, Shaanxi, China}
\affiliation{KEK Theory Center and Sokendai, Tsukuba, Ibaraki 305-0801, Japan}
\affiliation{Department of Particle and Nuclear Physics,
The Graduate University for Advanced Studies (Sokendai), Tsukuba 305-0801, Japan}

\author{Shingo Mori}
\email[Electronic address: ]{smori@post.kek.jp}
\affiliation{Department of Particle and Nuclear Physics,
The Graduate University for Advanced Studies (Sokendai), Tsukuba 305-0801, Japan}

\date{\today}

\begin{abstract}
We propose a novel method to reconstruct event by event the full kinematics of the cascade decay process, $h \to \tau^+\tau^- \to (\pi^+ \bar{\nu}) (\pi^- \nu)$, which allows us to measure the $\tau^+\tau^-$ spin correlation, a measure of the CP property of the Higgs boson. By noting that the $\tau^{\pm}$ momenta lie on the plane spanned by the accurately measured impact parameter and momentum vectors of charged pions, we can obtain the most likely momenta of the two missing neutrinos by using the probability distribution functions of the $\met$ vector and the location of the primary vertex.  A simple detector level simulation shows an excellent agreement between the reconstructed and the true kinematics, both in the $\tau^+\tau^-$ and the $\pi^+\pi^-$ rest frames.  The method can be tested in $Z \to \tau^+\tau^-$ events, which should exhibit no correlation.
%
\end{abstract}


\maketitle


%
\setcounter{page}{1}
\renewcommand{\thefootnote}{\arabic{footnote}}
\setcounter{footnote}{0}

CP property of the observed Higgs particle $h(125)$~\cite{ATLAS:Higgs2012, CMS:Higgs2012} is a window of the physics of mass generation.
In general the mass eigenstate $h(125)$ can be a mixture of CP-even and CP-odd scalar particles.
While only one CP-even scalar particle exists in the Standard Model (SM), many of its extensions not only modify the Higgs couplings to gauge bosons and fermions, but also predict additional scalars and pseudo-scalars.
If the Higgs sector is CP conserving, all the neutral mass eigenstates should have definite CP parity.
The pure CP eigenstate assumption has been investigated experimentally by both ATLAS and CMS collaborations~\cite{ATLAS:Higgs:CP2013, CMS:Higgs:CP2013, CMS:Higgs:CP4l2014}, and the CP-odd hypothesis is disfavored by nearly $3\sigma$. 

However, if the $h(125)$ particle is a mixture of the CP-even and CP-odd states, the bound on the mixing parameter is rather weak and a large mixing in the Higgs sector is still allowed~\cite{Brod:2013,Shu:2013,Dolan:2014}.
(For the recent review see~\cite{Heinemeyer:2013tqa} and references there in.)
There are several channels that can be used to measure the CP property of $h(125)$.
The golden channel $h\to ZZ^{\ast}/Z\gamma^{\ast}/\gamma^{\ast}\gamma^{\ast} \to (\ell\bar{\ell})(\ell'\bar{\ell}')$ has been analyzed in Refs.~\cite{Chen:2014-1, Chen:2014-2, Bishara:2014,Korchin:2013,Bernreuther:2010}.
The sensitivity is rather low because of the dominance of the tree-level (CP-even) $hZZ^{\star}$ amplitudes and the small (loop suppressed) $hZ\gamma^{\ast}$ and $h\gamma^{\ast}\gamma^{\ast}$ amplitudes.
Processes $pp\to hjj$~\cite{hjj}, $pp\to h t \bar{t}$~\cite{ttbar}, and $h \to \tau^{+}\tau^{-}$~\cite{htautau, He:1994} have also been analyzed. \draftnote{
In Ref.~\cite{Bower:2002}, it was pointed out that the correlation between planes spanned by $\pi^{\pm}$ and $\pi^0$ from the $\tau^{\pm}\to\rho^{\pm}\nu_\tau\to\pi^{\pm}\pi^0\nu_\tau$ decays can be used to measure CP violation,
and the experimental sensitivity can be improved by using the impact parameters~\cite{Desch:2003}. 
Alternatively, without using of the impact parameter, reconstruction of the internal substructure of those decay modes 
can also enhance the sensitivity~\cite{Harnik:2013}.
}


In Ref.~\cite{Berge:2008}, the 3-prong decay mode of tau was proposed to measure CP violation, for which
the tau momentum direction can be reconstructed directly, but the sensitivity is low, because of small 3-prong decay rate and the necessary spin projection to the longitudinal polarized state.
In Refs.~\cite{Berge:2009,Berge:2011,Berge:2013}, a new observable made of the impact parameters and the momenta of charged decay products was proposed.

In this letter we report our study on the process $pp \to h \to \tau^{+}(\pi^{+} \bar{\nu}_{\tau})\tau^{-}(\pi^{-} \nu_{\tau})$, in which the impact parameter vectors of the $\pi^{+}$ and $\pi^{-}$ in $\tau^{+}$ and $\tau^-$ decays are used to reconstruct event by event the full kinematics. 

In the analysis below we assume for simplicity the measured \revision{Higgs particle $h(125)$} is a mixture of CP-even and CP-odd scalars, denoted by $H$ and $A$ respectively,
\bee\label{eq:higgs:mix}
\revision{h = \cos\xi\, H + \sin\xi\, A}\,,
\ene
\revision{where $\xi$ is the Higgs mixing angle that has been assumed to be real.} We also assume the Yukawa interactions of $H$ and $A$ with tau-lepton pair are CP conserving, 
\bee
\mathcal{L} = - g_{H\tau\tau} H \bar{\tau} \tau - i  g_{A\tau\tau} A \bar{\tau}  \gamma^{5} \tau\,,
\ene
such that the only source of CP violation is in the mixing \iref{eq:higgs:mix}. The interactions between the mass eigenstate $h(125)$ and the tau-lepton pair are then described as
\bee
\revision{\mathcal{L} = - g_{h\tau\tau} h \big( \cos \xi_{h\tau\tau} \bar{\tau} \tau + i  \sin\xi_{h\tau\tau} \bar{\tau}  \gamma^{5} \tau\big)}\,,
\ene
where 
\begin{eqnarray}
 g_{h\tau\tau} &=& \sqrt{ (g_{H\tau\tau}\cos \xi)^2 + (g_{A\tau\tau}\sin \xi)^2},\\
  \xi_{h\tau\tau} &=& \tan^{-1}\left[(g_{A\tau\tau} /g_{H\tau\tau})\tan \xi\right], 
\end{eqnarray}
are, respectively, the magnitude and the CP-odd phase of the $h\tau\bar\tau$ coupling.
\draftnote{
Although the CP-violating interactions alter the branching ratios. However, we use in this report the SM branching ratio of $B(h\to\tau^{+}\tau^{-}) = 6.1\%$~\cite{PDG} to estimate the experimental sensitivity. 
It was shown in Ref.~\cite{Harnik:2013} that experimental sensitivity of  $\Delta\xi_{h\tau\tau}$ 
is about $0.2$ for LHC14 with an integrated luminosity $3\ab^{-1}$. The sensitivity can reach $0.05$ 
for ILC at $\sqrt{s}=500\gev$ with $1\ab^{-1}$~\cite{Berge:2013}. 
}

In our approximation of neglecting potential CP violation in $\tau$ decays, the CP-odd spin correlation of $\tau^+$ and $\tau^-$ can be measured by studying their decay correlations.
One of the observables with maximum sensitivity to the spin correlation is the azimuthal angle correlation in the Higgs rest frame, which has a simple form,
\bee\label{eq:corr}
\frac{1}{\Gamma} \frac{ d \Gamma }{ d \phi } 
= \frac{1}{2\pi} \bigg( 1 - \frac{\pi^2}{16 } \cos( \phi - 2 \xi_{h\tau\tau} ) \bigg)\,,
\ene
in the $m_\tau^2/m_h^2\to 0$ limit, 
where $\phi$ is the azimuthal angle of $\pi^{-}$ about the $\tau^{-}$ momentum as the $z$-axis, when the $x$-axis is chosen along the $\pi^{+}$ transverse momentum.
Exactly the same distribution~\iref{eq:corr} is found for the azimuthal angle $\phi$ of $\tau^-$ momentum
in the $\pi^{+}\pi^{-}$ rest frame, where the $z$-axis is along the $\pi^-$ momentum and the $x$-axis is along the $\tau^+$ transverse momentum. 
The advantage of the latter frame is that the $z$-axis can be directly reconstructed by the accurately measured $\pi^{+}$ and $\pi^-$ momenta.
In both frames, we should determine the $\tau^\pm$ momenta $\vec{p}_{\tau^{\pm}}$ accurately.


If the $\pi^\pm$ momenta are measured accurately, two parameters of the $\tau^\pm$ momenta can be determined by using the on-shell conditions.
We take remaining four parameters as the magnitude of the momentum vector of taus, $|\vec{p}_{\tau^{\pm}}|$, and the azimuthal angle of the taus, $\phi_{\tau^{\pm}}$, in the lab frame where the pion momentum, $\vec{p}_{\pi^{\pm}}$, is along the $z$(polar)-axis and the $x(p_{\tau^\pm}^x=0)$-axis in the scattering plane spanned by the beam and the $\pi^\pm$ momenta.

If we constrain the sum of the transverse momenta of the two neutrinos by the observed missing transverse momentum, the most likely values of $p_{\tau^{\pm}}$ distributes around their true values, allowing us to estimate the invariant mass of tau pair, $m_{\tau\tau}\simeq 2 |\vec{p}_{\tau^-}| |\vec{p}_{\tau^+}|(1-\cos(\theta_{\pi^+\pi^-}))$ in the collinear approximation~\cite{Rainwater:1998kj}.
However, the optimal values of the azimuthal angle, $\phi_{\tau^{\pm}}$, show virtually no correlation with their true values~\cite{preprint}. The azimuthal angle correlation~\iref{eq:corr} in the $\pi^+\pi^-$ rest frame is smeared out.

Fortunately, the $\tau$'s from Higgs decay have large decay lengths $|\vec{l}_{\tau^{\pm}}|$, typically of $c\tau_\tau(m_h/2m_\tau)\sim3.1~\mm$. 
Therefore the impact parameter vectors $\vec{b}_{\pi^{\pm}}$ of $\pi^{\pm}$ can be measured with a significant efficiency, providing us with the desired construction of the azimuthal angle, $\phi_{\tau^\pm}$, in the lab frame.

For single tau decay, $\tau^{-}\to\pi^{-}\nu_{\tau}$, once the impact parameter vector $\vec{b}_{\pi^{-}}$ is measured, the decay plane is accurately determined by $\vec{b}_{\pi^-}$ and $\vec{p}_{\pi^-}$, which are orthogonal $\vec{b}_{\pi^-}\cdot\vec{p}_{\pi^-}=0$.
The $\tau$ momentum $\vec{p}_{\tau^-}$ should lie on this plane and the opening angle between $\vec{p}_{\tau^-}$ and $\vec{p}_{\pi^-}$ is constrained by the on-shell condition
\bee
\cos\theta_{\tau^{-}\pi^{-}} 
= 
\frac{ 2   E_{\tau^{-}}   E_{\pi^{-}} - m_{\tau}^2 - m_{\pi^{-}}^2 }
{ 2 |\vec{p}_{\tau^{-}}| |\vec{p}_{\pi^{-}}| }\,,
\ene
where $|\vec{p}_{\tau^-}|$ is the only unknown.
The orientation of the $\tau^{-}$ momentum can be solved directly
\bee\label{eq:ptau}
\frac{\vec{p}_{\tau^{-}} }{ |\vec{p}_{\tau^{-}}| } = \frac{ \vec{b}_{\pi^{-}} + \frac{ |\vec{b}_{\pi^{-}}| } { \tan\theta_{\tau^{-}\pi^{-}} } \frac{\vec{p}_{\pi^{-}} }{ |\vec{p}_{\pi^{-}}| } } { \bigg|\vec{b}_{\pi^{-}} + \frac{ |\vec{b}_{\pi^{-}}| } { \tan\theta_{\tau^{-}\pi^{-}} } \frac{\vec{p}_{\pi^{-}} }{ |\vec{p}_{\pi^{-}}| } \bigg| }\,,
\ene
where the sign of the second term is fixed by the condition
$
(\vec{p}_{\tau^{-}} \cdot \vec{p}_{\pi^{-}}) > 0\,.
$
The same applies for $\tau^+\to\pi^+\bar{\nu}_\tau$ decay, leaving only two free parameters $|\vec{p}_{\tau^{-}}|=p_{\tau^-}$ and $|\vec{p}_{\tau^{+}}|=p_{\tau^+}$ to reconstruct the full kinematics of the process.

It is at this stage we impose the $\met$ constraint with the probability distribution function (PDF),
\begin{equation}
\rho_{\met}(p_{\tau^{\pm}})
=
\frac{1}{\cal N} \exp\bigg[ 
-\frac{1}{2}
( \Delta\met(p_{\tau^{\pm}})  )^{T} V^{-1} ( \Delta\met(p_{\tau^{\pm}}) )
\bigg],\label{eq:density}
\end{equation}
\begin{equation}
 V = R(\phi_{\met})
  \begin{pmatrix}
   \sigma_{\met}^2 & 0\\
   0 & |\met^{\, \rm obs}|^2 \sigma_{\phi_{\met}}
  \end{pmatrix}
  R^{-1}(\phi_{\met}),  
\end{equation}
where ${\cal N} = 2\pi \sqrt{\det V}$, for the $\Delta\met(p_{\tau^{\pm}}) = \met(p_{\tau^{\pm}}) - \met^{\,\rm obs}$ is the difference between the observed and the expected $\met$ vectors, $R(\phi)$ is the rotation about the beam $(z)$-axis.
This PDF measures the likelihood that the observed $\met^{\, \rm obs}$ is compatible with the sum of $\vec p_{\nu} + \vec p_{\bar \nu}$, which is a function of $p_{\tau^\pm}$.
Here, the $\met$ resolution is represented by the covariance matrix $V$, which is, in principle, estimated on an event-by-event basis in the detector-level simulation, following the algorithm of~\cite{CMS:MET}.




Below we explain how we simulate the process $pp \to h \to \tau^{+}\tau^{-} \to (\pi^{+} \bar{\nu}_{\tau})(\pi^{-} \nu_{\tau})$.
The events are generated at LO for $\sqrt{s} = 14~\tev$ by using MadGraph5~\cite{madgraph5}.
The Higgs production process is simulated by the HC model file~\cite{mg5:model:hc}, and the $\tau^+\tau^-$ spin correlation is obtained by using the TauDecay package~\cite{Hagiwara:2013}.
The generated events are then showered by Pythia8~\cite{Pythia8}, and the detector effects are simulated by using Delphes3~\cite{Delphes3}. The jets are classified by using the FastJet package~\cite{FastJet} with anti-$k_{T}$ algorithm and a distance $\Delta R = 0.4$. 

The $\tau$-jets are tagged by using the Delphes3 algorithm which has a reconstruction efficiency of about $0.8$ for signal and $0.6$ for $Z\to\tau^+\tau^-$ events.
We multiply this efficiency by the $\tau$-identification efficiency which is about $0.6$ for a medium tau-jet identification condition and has a fake rate about $1\%$ from QCD jets~\cite{ATLAS:HiggsTauTau,CMS:HiggsTauTau}. 

The directions of $\pi^{\pm}$ momenta are chosen as the exact in first, and then smeared by using the current resolutions of tracks~\cite{ATLAS:InnerD:2010}.
The magnitudes of $\pi^{\pm}$ momenta are smeared to be the corresponding $\tau$-tagged jets momentum.
Using tracks inside of the $\tau$-tagged jets is essential because the soft particles inside of the $\tau^{\pm}$-tagged jets could completely wash out the relative orientation between $\tau^{\pm}$ and $\pi^{\pm}$.

The observed missing transverse momentum $\met^{\, \rm obs}$ is calculated on an event-by-event basis by using the Delphes3.
We determine the resolution of the missing transverse momentum $\sigma_{\met}$ and its azimuthal angle $\sigma_{\phi_{\met}}$ by comparing the sum of neutrino momenta at parton level with the observed $\met$ and $\phi_{\met}$ at detector level.
We have checked these errors are consistent with those calculated from errors of all visible tracks~\cite{preprint}.

The exact impact parameter vectors $\vec{b}_{\pi^{\pm}}$ are derived using exact decay length vectors of tau $\vec{l}_{\tau^{\pm}}$ given by Pythia8.
For those events with $p_\tau\sim m_h/2$, we find the impact parameter distribution to be exponentially falling with the mean of $|\vec{b}_{\pi^{\pm}}|\sim100~\mum$.
In practice, the location of the primary vertex is not known accurately~\cite{ATLAS:InnerD:2010}, and we should compute the impact parameter vectors from the most likely location at the primary vertex.
Although the error might be smaller for those events with two isolated $\pi^+$ and $\pi^-$ trajectories that we study, we introduce a Gaussian smearing distribution with resolutions $\sigma_{b_{T}} = 20~\mum$ and $\sigma_{b_{Z}} = 40~\mum$ in the transverse and in the beam directions~\cite{ATLAS:InnerD:2010}.
Therefore, we obtain the smeared impact parameter vectors $\vec{b}_{\pi^{\pm}}^{\, \rm obs}$ from exact decay length vectors $\vec{l}_{\tau^{\pm}}$ and the smeared primary vertex.


\draftnote{For background, we consider here only the dominant irreducible process, $pp\to Z\to \tau\tau$.
Fake backgrounds from QCD jets may also contribute. It is shown in Ref.~\cite{ATLAS:HiggsTauTau}  that at $\sqrt{s}=7, 8\tev$ about $21\%$ (gluon fusion dominated region) and $42\%$ (VBF dominated region) of the total background are fake backgrounds. At $\sqrt{s}=14\tev$, those values may grow, but since we employ only the double single $\pi$ decay modes and since the fake background does not give azimuthal angle correlation, we believe that our estimate based on $Z\to \tau\tau$ is valid especially after improving the impact parameter cuts. 
}
The efficiencies and number of events are summarized in Table~\ref{tab:events}.
The production cross sections of the signal $\sigma(pp\to h+{\rm anything})=62.1~\pb$~\cite{Heinemeyer:2013tqa} and the background $\sigma(pp\to Z+{\rm anything})=62.2~\nb$~\cite{Catani:2009sm} give about $1.2\times10^5$ and $7.0\times10^7$ events for the signal $(h\to\tau^+\tau^-\to\pi^+\pi^-\nu\bar\nu)$ and the background $(Z\to \tau^+\tau^-\to \pi^+\pi^-\nu \bar \nu)$ at $3~\ab^{-1}$, respectively.
 The double tau-tag efficiency is about $(0.8\times0.6)^2\sim0.2$ for signal and $(0.6\times0.6)^2\sim0.1$ for the background events.
 Our final state selection cuts on $|\vec{p}_{\pi^{\pm}, T}^{\rm \ obs}|$, $|\eta_{\pi^{\pm}}^{\rm obs}|$, and $|\met^{\rm \ obs}|$ reduce the events by a factor of $0.18$ for the signal and $0.01$ for the background.
 

It is these selected events we find the most likely values of $p_{\tau^-}$ and $p_{\tau^+}$ by using the smeared $\vec b_{\pi^\pm}$ vectors and the PDF~\iref{eq:density} of the missing $p_T$ vector.
The method gives a good resolution for the invariant mass of the $\tau$-pair~\cite{preprint}, 
\draftnote{(other method for the mass reconstruction can be found in Refs. 
\cite{ATLAS:HiggsTauTau, CMS:HiggsTauTau, Konar:2016})},
and we impose $|m_{\tau\tau}^{\rm obs}- m_h|<10~\gev$. 
We find that $0.49$ of signal survives the cut, while the background is suppressed by $0.075$.

\begin{table}[h]
\vspace{-0.2cm}
\caption{Efficiency and expected number of events for the signal process $pp \to h \to  \tau^{+}\tau^{-} \to (\pi^{+} \bar{\nu}_{\tau})(\pi^{-} \nu_{\tau})$, and the major irreducible background process  $pp \to Z \to  \tau^{+}\tau^{-} \to (\pi^{+} \bar{\nu}_{\tau})(\pi^{-} \nu_{\tau})$, at $14~\tev$ with an integrated luminosity $3~{\rm ab^{-1}}$. }
\vspace{-0.3cm}
\begin{center}
\begin{tabular}{c||c|c||c|c}
\hline\hline
 & {\rm Eff. } & {\rm Evt.($h$) } & {\rm Eff. } & {\rm Evt.($Z$) } 
\\\hline
{\rm No cuts}  
&  $ 1.000 $ &  $ 1.42\times10^{5} $ 
	 &  $ 1.000 $ &  $ 7.31\times10^{7} $

\\\hline
{\rm tau-tag}  
&  $ 0.225 $ &  $ 3.18\times10^{4}$ 
&  $ 0.120 $ &  $ 8.78\times10^{6}$

\\\hline
\begin{tabular}{c}
$|\eta_{\pi^{\pm}}^{\rm obs}|<2.5$
\\
${\rm min}(|\vec{p}_{\pi^{\pm}, T}^{\rm \ obs}|) > 15~\gev $  
\\
${\rm max}(|\vec{p}_{\pi^{\pm}, T}^{\rm \ obs}|) > 35~\gev $      
\\
$|\met| > 45~\gev $
\end{tabular} 
& $0.180$ & $5.72\times10^{3}$ 
& $0.010$ & $8.78\times10^{4}$
		 		 \\\hline
$|m_{\tau\tau}-m_h| < 10~\gev$& 
$ 0.492 $ & $2.81\times10^{3}$ & 
$ 0.075 $ & $6.58\times10^{3}$
\\\hline\hline
$\min(|\vec{b}_{\pi^{\pm},T}^{\rm \ obs}|) > 50~\mum$ 
& $0.150$ & $422$ 
& $0.240$ & $1.58\times10^{3}$

\\\hline\hline
\end{tabular}
\end{center}\label{tab:events}
\vspace{-0.5cm}
\end{table}%


\begin{figure}[t]
\vspace{-0.2cm}
\includegraphics[scale=0.4,angle=0]{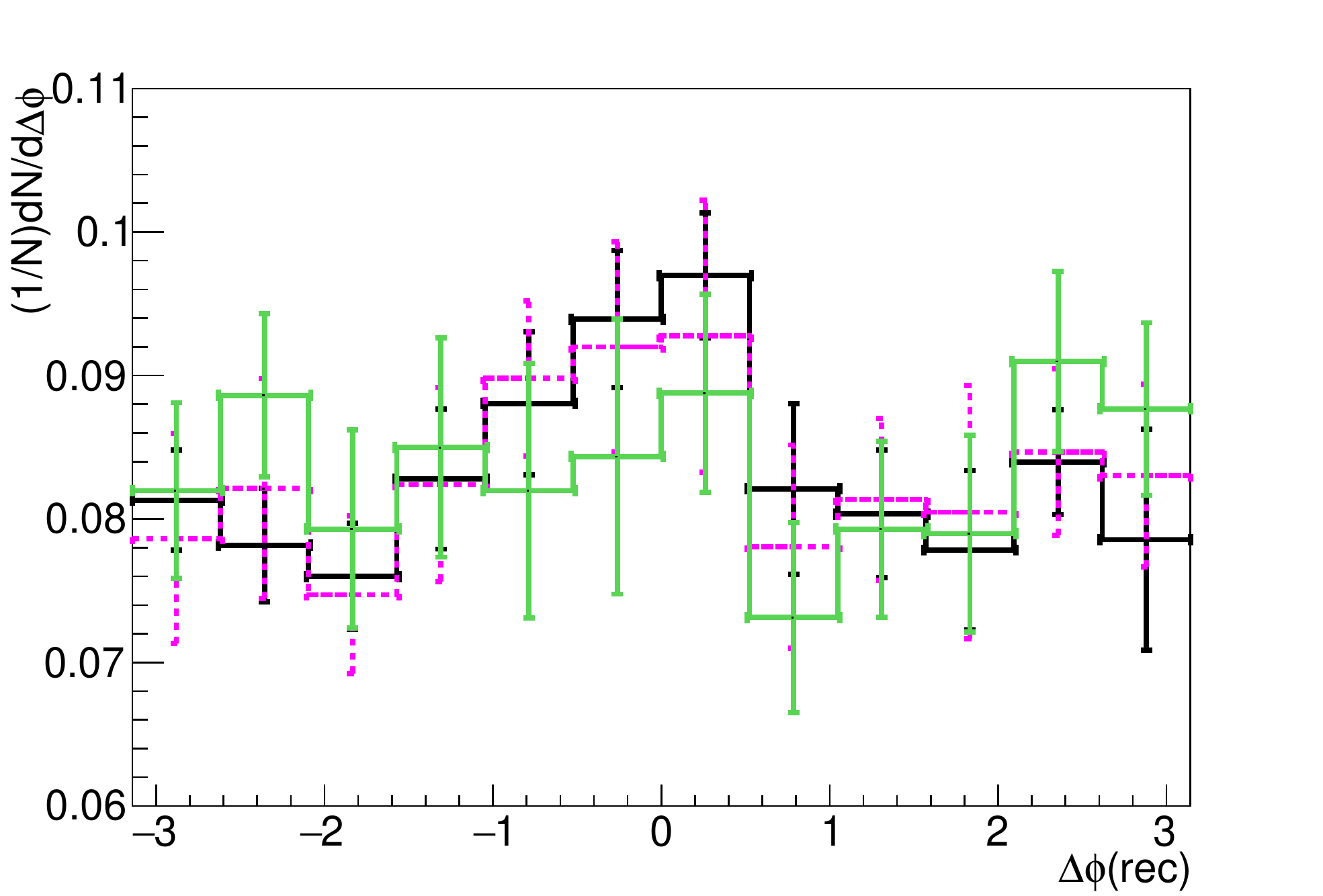}
\vspace{-0.3cm}
\caption{The reconstructed azimuthal angle distribution of $Z\to\tau^+\tau^-$ events after the smearing in the impact parameter. The black-solid line denotes the case without $|\vec b_{\pi^{\pm},T}^{\rm \ obs}|$ cut, the pink-dotted line denotes $\min(|\vec{b}_{\pi^{\pm},T}^{\rm \ obs}|)>25~\mum$, and the green-solid line shows $\min(|\vec{b}_{\pi^{\pm},T}^{\rm \ obs}|)>50~\mum$. The data points correspond to an integrated luminosity $3~\ab^{-1}$.}
\vspace{-0.5cm} 
\label{fig:bias}
\end{figure}

In Fig.~\ref{fig:bias}, the normalized $\Delta\phi^{\rm rec}$ distribution of $Z\to\tau^+\tau^-\to \pi^+\pi^-\nu \bar \nu$ events is shown, where the black-solid line is obtained after the smearing in the impact parameter vector is introduced.
The pink-dotted line is found after imposing the cut $|\vec b_{\pi^\pm,T}^{\rm \ obs}|>25~\mum$, and the green-solid line for $|\vec b_{\pi^\pm,T}^{\rm \ obs}|>50~\mum$.
After the last cut the distribution becomes flat as the theoretical prediction. 
This is because those events whose true $|\vec{b}_{\pi^\pm}|$ are smaller than experimental resolution cannot be resolved, and our reconstruction procedure via Eq.~\iref{eq:ptau} tends to give $\Delta\phi^{\rm rec}\sim0$ for all those events with $|\vec{b}_{\pi^\pm}^{\rm \ true}|\ll|\vec{b}_{\pi^\pm}^{\rm \ obs}|$~\cite{preprint}.
Fortunately, as is shown in Fig.~\ref{fig:bias} this systematic bias can be reduced by applying cuts on $|\vec{b}_{T}^{\rm \ obs}|$.
As shown in the bottom line of Tab.~\ref{tab:events}, the efficiency of impact-parameters cut for the Higgs decay $0.15$ is smaller than the one for $Z\to\tau^+\tau^-\to \pi^+\pi^-\nu \bar \nu$ events $0.24$ at $14~\tev$ because the momentum of the softer $\pi^\pm$ is lower for the signal than the background after the $|m_{\tau\tau}^{\rm obs}- m_h|<10~\gev$ cut, due to the chirality flipping nature of the $h\tau\bar\tau$ coupling~\cite{preprint}.
In the end, we find $S/\sqrt{S+B} \approx 9.4$.
\begin{figure}[t]
\vspace{-0.2cm}
\includegraphics[scale=0.4,angle=0,origin=c]{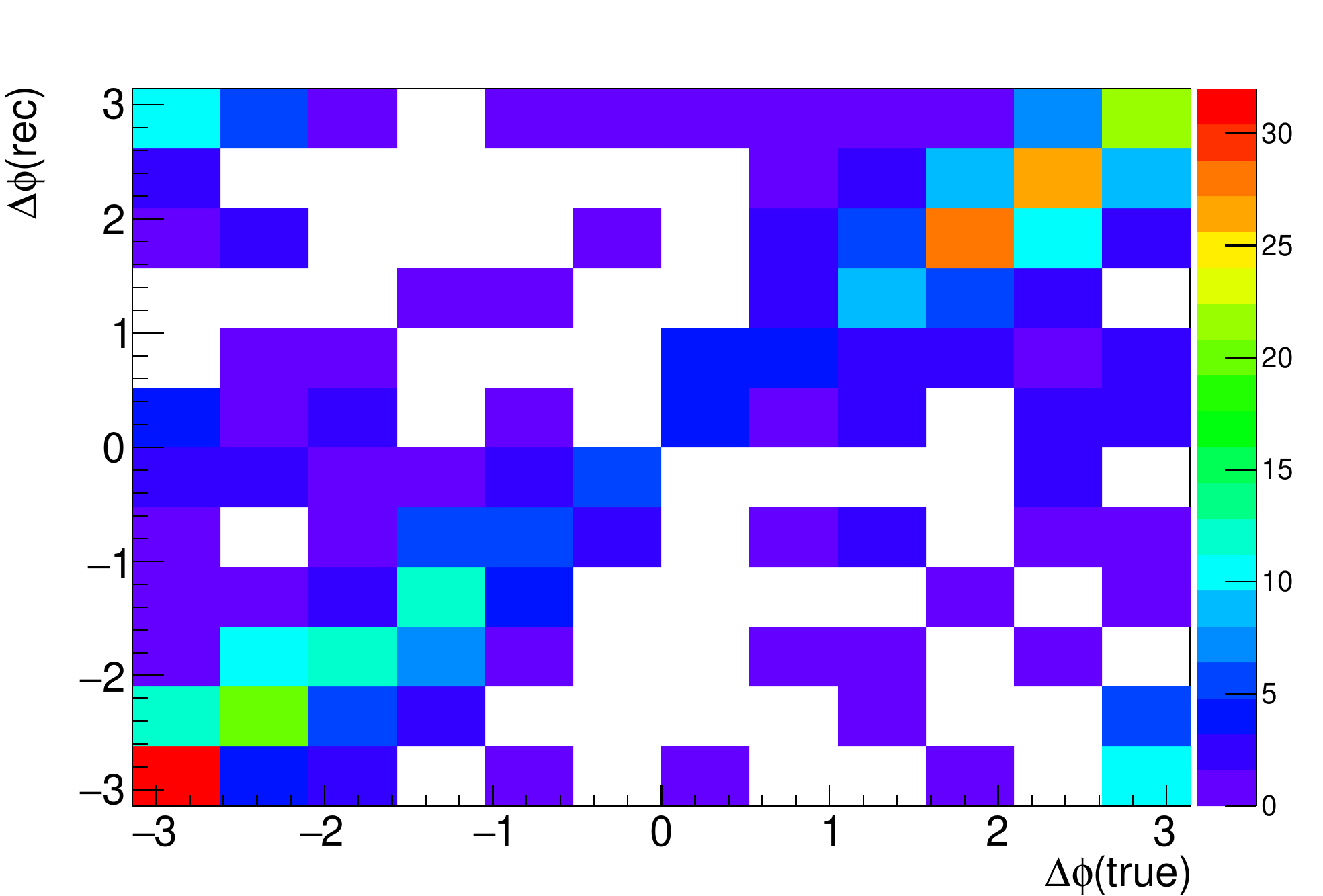}
\vspace{-0.3cm}
\caption{Correlations between the true and reconstructed azimuthal angle difference for the SM Higgs $(\xi_{h\tau\tau}=0)$ after the cut $|\vec b_{\pi^\pm,T}^{\rm \ obs}|>50~\mum$. The $422$ data points correspond to an integrated luminosity 3~$\rm ab^{-1}$.}
\vspace{-0.5cm}
\label{fig:a}
\end{figure}
\begin{figure}[b]
\vspace{-0.5cm}
\centering\includegraphics[scale=0.4,angle=0]{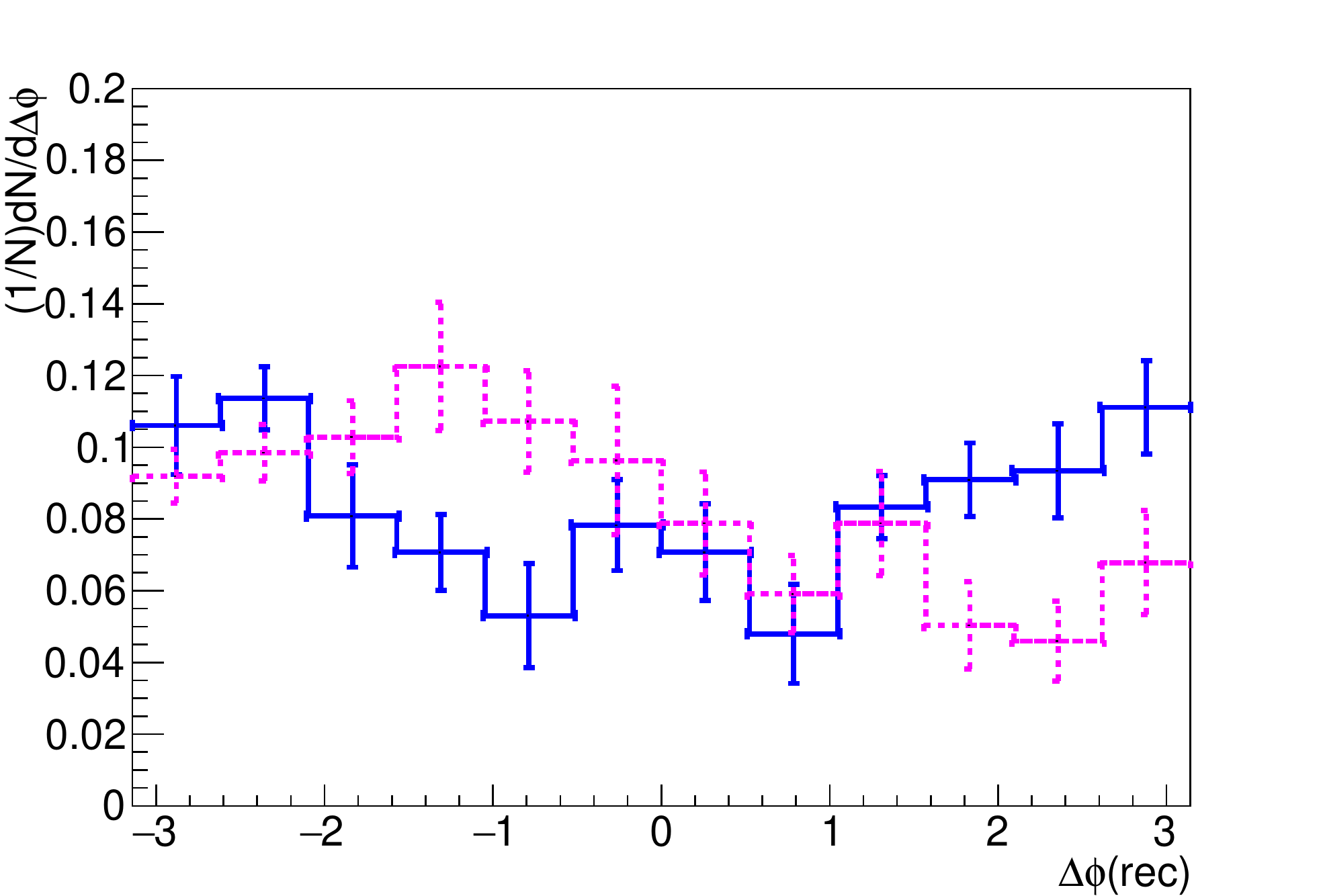}
\vspace{-0.3cm}
\caption{Distributions of the reconstructed azimuthal angle difference for the $h\to \tau^+\tau^-\to \pi^+\pi^- \nu \bar \nu$ events with $\xi_{h\tau\tau}=0$ (blue-solid line) and $\xi_{h\tau\tau}=\pi/4$ (pink-dashed line) after the cut $|\vec b_{\pi^\pm,T}^{\rm \ obs}|>50~\mum$. The data points correspond to an integrated luminosity 3~$\rm ab^{-1}$.}
\vspace{-0.5cm}
\label{fig:b} 
\end{figure}


Fig.~\ref{fig:a} shows the correlation in the $\pi^{+}\pi^{-}$ rest frame between the true and reconstructed azimuthal angle for the SM Higgs boson, {\it i.e.} $\xi_{h\tau\tau}=0$, after the cut $|\vec b_{\pi^\pm,T}^{\rm \ obs}|>50~\mum$.
The reconstructed $\Delta\phi$ distributes around the true value within about $\pi/6$ accuracy for all $\Delta\phi$(true) values.
We find that the $\Delta\phi({\rm rec})\mathchar`-\Delta\phi({\rm true})$ agreement is worse~\cite{preprint} in the $\tau^{+}\tau^{-}$ rest frame, because the reconstructed $\tau^\pm$ momenta have relatively larger error.
We, therefore, propose to use the $\pi^{+}\pi^{-}$ rest frame to study the decay plane correlation. 
Shown in Fig.~\ref{fig:b} are the reconstructed $\Delta\phi$ distribution of the signal events for the SM $(\xi_{h\tau\tau}=0)$ in blue-solid and for maximum CP violation $(\xi_{h\tau\tau}=\pi/4)$ in pink-dashed lines, after the cut $|\vec b_{\pi^\pm,T}^{\rm \ obs}|>50~\mum$ is applied.
We can measure clearly CP violation as a phase shift in the $\Delta\phi$ distribution~\iref{eq:corr}, if the background is absent.

\begin{figure}[t]
\vspace{-0.2cm}
\includegraphics[scale=0.4,angle=0]{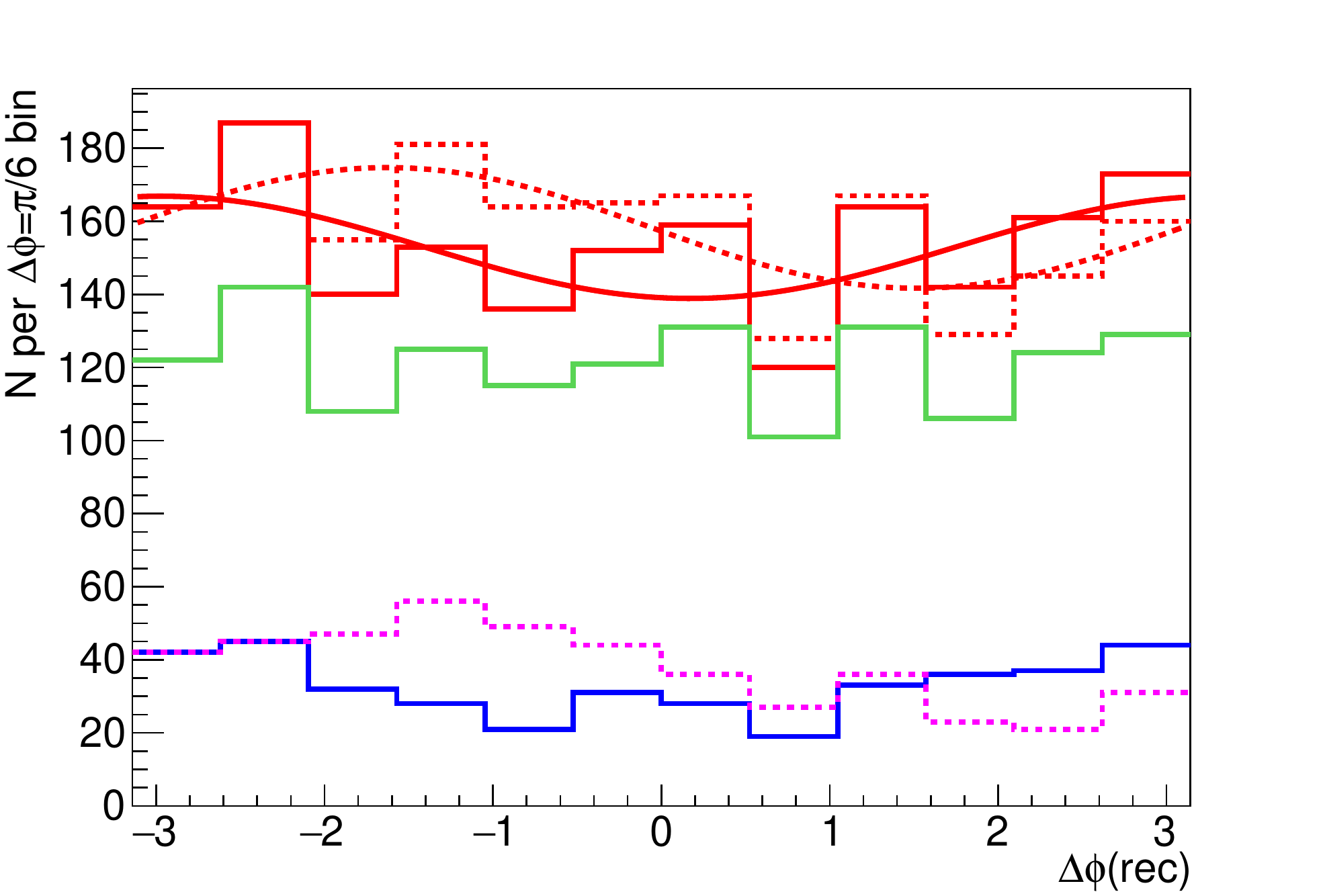}
\vspace{-0.3cm}
\caption{The $\Delta \phi^{\rm rec}$ distribution of the signal and background and the result of fitting. The blue-solid line show the signal events of $\xi_{h\tau\tau}=0$, the green-solid line shows the background events. The red-solid histogram shows their sum. The red-solid curve shows our fit. The dashed line and histograms are for $\xi_{h\tau\tau}=\pi/4$. In both cases, we use the same background events. We require $|\vec{b}_{\pi^{\mp},T}^{\rm \ obs}| > 50~\mum$. The data points correspond to an integrated luminosity $3~\ab^{-1}$.}
\vspace{-0.5cm}
\label{fig:bkg}
 \end{figure}

Fig.~\ref{fig:bkg} shows histograms of $\Delta\phi^{\rm rec}$ for signal, background and their sum after the cut $|\vec{b}_{\pi^{\mp},T}^{\rm \ obs}| > 50~\mum$.
The blue-solid and pink-dashed lines denote the signal events for $\xi_{h\tau\tau}=0$ and $\pi/4$, respectively.
The green-solid line shows the background events.
The red-solid and -dotted curves show our fit to the sum of background and signal events for $\xi_{h\tau\tau}=0$ and $\xi_{h\tau\tau}=\pi/4$, respectively.
The fit function is simply the sum of the function (\ref{eq:corr}) and the constant background, where their normalizations and the phase shift, $\xi_{h\tau\tau}$ in Eq.~\iref{eq:corr}, are fitted to the binned data as shown by the red histograms in Fig.~\ref{fig:bkg}.
We find for $\xi_{h\tau\tau}^{\rm true}=0$ and $\pi/4$, respectively, $\xi_{h\tau\tau}=0.030\pm0.19$ at $\chi_{\rm min}^2/{\rm d.o.f}=14.8/9$, and $\xi_{h\tau\tau}=0.78\pm 0.18$ at $\chi_{\rm min}^2/{\rm d.o.f}=13.6/9$.
We checked the result is stable under the change of bin size.



The sensitivity of $\Delta\xi_{h\tau\tau}\approx0.2$ from the $\tau^+\tau^-\to \pi^+\pi^-\nu\bar \nu$ mode only is encouraging.
And what is more, we find that the kinematical correlation as shown in Fig.~\ref{fig:bias} can be parametrized as a function of the cut-off parameter, $\min(|\vec b_{\pi^\pm}^{\rm \ obs}|)$.
By modifying the fitting function to account for the kinematical bias, we find significant improvements in the $\Delta\xi_{h\tau\tau}$ accuracy of  possibly a factor of $10$, details of which will be reported elsewhere~\cite{preprint}.
We believe that the method can be tested and improved by using the side bands, {\it e.g.} for those events which satisfy $|m_{\tau\tau}^{\rm obs}- m_Z|<10~\gev$ or $m^{\rm obs}_{\tau\tau}>150~\gev$, which are dominated by $Z\to \tau^+\tau^-$ background. 

\revision{In summary, by employing the impact parameter vectors of $\pi^\pm$ trajectories, we propose a novel method to measure the CP violation in $h \to \tau^{+}\tau^{-}\to \pi^+\pi^-\nu\bar\nu$. Even through only part of the kinematical information of tau leptons is stored in the $\pi^\pm$ momenta and the impact parameters, $\vec p_{\pi^\pm}$ and $\vec b_{\pi^\pm}$, the spin correlation can still be measured by maximizing the probability densities, \eq{eq:density}, for the missing transverse momenta, $\met$.
We find an excellent agreement between the reconstructed and true kinematics in the $\pi^{+} \pi^{-}$ rest frame, by using the typical experimental resolutions of the LHC detectors. The experimental sensitivity is estimated to be $\Delta\xi_{h\tau\tau} \approx 0.2$ with an integrated luminosity $ 3~\mathrm{ab}^{-1}$ at $\sqrt{s}=14~\tev$.}


\section*{Acknowledgements}
K.H. is supported in part by the William F. Vilas Trust Estate, and by the U.S. Department of Energy under the contract DE-FG02-95ER40896. 
K.M. is supported by the China Scholarship Council, and the National Natural Science Foundation of China under Grant No. 11647018, and partially by the Project of Science and Technology Department of Shaanxi Province under Grant No. 15JK1150.



\begin{thebibliography}{99}

\bibitem{ATLAS:Higgs2012}
G. Aad \etal. (ATLAS Collaboration), 
Phys. Lett. {\bf B716}, 1(2012). 

\bibitem{CMS:Higgs2012}
S. Chatrchyan \etal (CMS Collaboration), 
Phys. Lett. {\bf B716}, 30 (2012).


\bibitem{ATLAS:Higgs:CP2013}
G. Aad \etal. (ATLAS Collaboration), 
Phys. Lett. {\bf B726}, 120(2013). 

\bibitem{CMS:Higgs:CP2013}
S. Chatrchyan \etal (CMS Collaboration), 
Phys. Rev. Lett. {\bf 110}, 081803(2013);
Erratum Phys. Rev. Lett. {\bf 110}, 189901(2013).

\bibitem{CMS:Higgs:CP4l2014}
S. Chatrchyan \etal (CMS Collaboration), 
Phys. Rev. {\bf D89}, 092007(2014).



\bibitem{Brod:2013}
J. Brod, U. Haisch, and J. Zupan, 
JHEP {\bf 1311}, 180 (2013).

\bibitem{Shu:2013}
J. Shu and Y. Zhang, 
Phys. Rev. Lett. {\bf 111}, 091801 (2013).

\bibitem{Dolan:2014}
M. J. Dolan, P. Harris, M. Jankowiak, M. Spannowsky,
Phys. Rev. {\bf D90}, 073008 (2014).


 \bibitem{Heinemeyer:2013tqa} 
  D. de Florian {\it et al.} [LHC Higgs Cross Section Working Group Collaboration],
  arXiv:1610.07922 [hep-ph].


\bibitem{Chen:2014-1}
Y. Chen, A. Falkowski, I. Low, R. Vega-Morales,
Phys. Rev. {\bf D90}, 113006(2014).

\bibitem{Chen:2014-2}
Y. Chen, R. Harnik, R. Vega--Morales,
Phys. Rev. Lett. {\bf 113}, 191801 (2014).

\bibitem{Bishara:2014}
F. Bishara, Y. Grossman, R. Harnik, D. J. Robinson, J. Shu and J. Zupan,
J. High Energy Phys. {\bf 04}, 084(2014).

\bibitem{Korchin:2013}
A. Y. Korchin, V. A. Kovalchuk,
Phys. Rev. {\bf D88}, 036009(2013).

\bibitem{Bernreuther:2010}
W. Bernreuther, P. Gonzalez, M. Wiebusch,
Eur. Phys. J. {\bf C69}, 31(2010).

 \bibitem{hjj}
	 F.~Campanario and M.~Kubocz, JHEP {\bf 1410}, 173 (2014).
	 C.~Englert, M.~Spannowsky and M.~Takeuchi, JHEP {\bf 1206}, 108 (2012).
	 J.~R.~Andersen, K.~Arnold and D.~Zeppenfeld, JHEP {\bf 1006}, 091 (2010).
	 G.~Klamke and D.~Zeppenfeld, arXiv:0705.2983 [hep-ph].
	 V.~Hankele, G.~Klamke and D.~Zeppenfeld, hep-ph/0605117.
	 T.~Plehn, D.~L.~Rainwater and D.~Zeppenfeld, Phys.\ Rev.\ Lett.\  {\bf 88}, 051801 (2002).


 \bibitem{ttbar}
	 N.~Mileo, K.~Kiers, A.~Szynkman, D.~Crane and E.~Gegner,arXiv:1603.03632 [hep-ph].
	 M.~R.~Buckley and D.~Goncalves, Phys.\ Rev.\ Lett.\  {\bf 116}, no. 9, 091801 (2016).
	 M.~Casolino, T.~Farooque, A.~Juste, T.~Liu and M.~Spannowsky, Eur.\ Phys.\ J.\ C {\bf 75}, 498 (2015).
	 F.~Boudjema, D.~Guadagnoli, R.~M.~Godbole and K.~A.~Mohan, Phys.\ Rev.\ D {\bf 92}, no. 1, 015019 (2015).
	 K.~Kołodziej and A.~Słapik, Eur.\ Phys.\ J.\ C {\bf 75}, no. 10, 475 (2015).
	 X.~G.~He, G.~N.~Li and Y.~J.~Zheng, Int.\ J.\ Mod.\ Phys.\ A {\bf 30}, no. 25, 1550156 (2015).
	 J.~Ellis, D.~S.~Hwang, K.~Sakurai and M.~Takeuchi, JHEP {\bf 1404}, 004 (2014).
	 F.~Demartin, F.~Maltoni, K.~Mawatari, B.~Page and M.~Zaro, Eur.\ Phys.\ J.\ C {\bf 74}, no. 9, 3065 (2014).
	 S.~Biswas, R.~Frederix, E.~Gabrielli and B.~Mele, JHEP {\bf 1407}, 020 (2014).
	 S.~Khatibi and M.~M.~Najafabadi, Phys.\ Rev.\ D {\bf 90}, no. 7, 074014 (2014).


 \bibitem{htautau}
	 J.~R.~Dell'Aquila and C.~A.~Nelson,
	 Nucl.\ Phys.\ B {\bf 320}, 61 (1989).
	 
 \bibitem{He:1994}
	 %
	 X.-G. He, J. P. Ma, B. McKellar, Mod. Phys. Lett. A {\bf9}, 205(1994).
	 A. Hayreter, X.-G. He, G. Valencia, Phys. Rev. D {\bf94}, 075002 (2016); Phys. Lett. B, {\bf760}, 175(2016).


\bibitem{Bower:2002}
	G. Bower, T. Pierzchala, Z. Was, and M. Worek, 
Phys. Lett. {\bf B543}, 227 (2002).

\bibitem{Desch:2003}
K. Desch, Z. Was, and M. Worek, 
Eur. Phys. J. {\bf C29}, 491 (2003).

\bibitem{Harnik:2013}
R. Harnik, A. Martin, T. Okui, R. Primulando, and F. Yu,
Phys. Rev. {\bf D88}, 076009 (2013).	

\bibitem{Berge:2008}
S. Berge, W. Bernreuther, and J. Ziethe, 
Phys. Rev. Lett. {\bf 100}, 171605 (2008).

\bibitem{Berge:2009}
S. Berge and W. Bernreuther, 
Phys. Lett. {\bf B671}, 470 (2009).

\bibitem{Berge:2011}
S. Berge, W. Bernreuther, B. Niepelt, and H. Spiesberger,
Phys. Rev. {\bf D84}, 116003 (2011).

\bibitem{Berge:2013}
S. Berge, W. Bernreuther, and H. Spiesberger,
Phys. Lett. {\bf B727}, 488(2013).





\bibitem{PDG}
K.A. Olive \etal, 
{\it (Particle Data Group)}, 
	Chin. Phys. {\bf C38}, 090001(2014).

\bibitem{Rainwater:1998kj} 
  D.~L.~Rainwater, D.~Zeppenfeld and K.~Hagiwara,
  Phys.\ Rev.\ D {\bf 59}, 014037 (1998).
	

\bibitem{CMS:MET}
CMS Collaboration, 
JINST, {\bf 6}, 09001(2011).

\bibitem{ATLAS:MMC}
A. Elagin \etal, 
Nucl. Instrum. Meth. {\bf A 654}, 481(2011).	

\bibitem{ATLAS:HiggsTauTau}
ATLAS Collaboration, 
	J. High Energy Phys. {\bf 04}, 117(2015).

	\bibitem{CMS:HiggsTauTau}
CMS Collaboration, 
J. High Energy Phys. {\bf05}, 104(2014).	

\bibitem{Konar:2016}
P. Konar, A. K. Swain,
Phys. Lett. B {\bf757}, 211(2016); Phys. Rev. D {\bf93}, 015021(2016). 
A. K. Swain, P. Konar, J. High Energy Phys. {\bf 03}, 142(2015). 
D. Jeans, NIM A, {\bf810}, 51(2016).





	
\bibitem{madgraph5}
J. Alwall, \etal,
J. High Energy Phys. {\bf 07}, 079(2014).


\bibitem{mg5:model:hc}
P. Artoisenet, P. Artoisenet, P. de Aquino, F. Demartin, R. Frederix \etal,
J. High Energy Phys. {\bf 11}, 043(2013).

\bibitem{Hagiwara:2013}
K. Hagiwara, T. Li, K. Mawatari, and J. Nakamura,
Eur. Phys. J. {\bf C73}, 2489(2013).

\bibitem{Pythia8}
T. Sj\"{o}strand, S. Mrenna and P. Skands, 
J. High Energy Phys., {\bf 05}, 026(2006), Comput. Phys. Comm. {\bf 178}, 852(2008).

\bibitem{Delphes3}
J. de Favereau, \etal,
J. High Energy Phys., {\bf 02}, 057(2014).


\bibitem{FastJet}
 M.~Cacciari, G.~P.~Salam and G.~Soyez,
Eur.\ Phys.\ J.\ C {\bf 72}, 1896(2012).


\bibitem{ATLAS:InnerD:2010}
ATLAS Collaboration, 
Eur. Phys. J. {\bf C70}, 787(2010).


\bibitem{Catani:2009sm} 
  S.~Catani, L.~Cieri, G.~Ferrera, D.~de Florian and M.~Grazzini,
  Phys.\ Rev.\ Lett.\  {\bf 103}, 082001 (2009).

	

\bibitem{preprint}
K.~Hagiwara, K.~Ma, and S.~Mori,
	(in preparation).



\end{thebibliography}
\end{document}